# Lived Experience in Dialogue: Co-designing Personalization in Large Language Models to Support Youth Mental Well-being

Kathleen W. Guan*

Delft University of Technology, k.guan@tudelft.nl, https://orcid.org/0000-0002-0044-0140

Sarthak Giri

University of Oulu, sarthak.giri@oulu.fi, https://orcid.org/0009-0008-5449-820X

Mohammed Amara

University of Oxford, mohammed.amara@univ.ox.ac.uk, https://orcid.org/0009-0001-0524-6830

Bernard J. Jansen

Hamad Bin Khalifa University, bjansen@hbku.edu.qa, https://orcid.org/0000-0002-6468-6609

Enrico Liscio

Delft University of Technology, e.liscio@tudelft.nl, https://orcid.org/0000-0002-8285-5867

Milena Esherick

Independent Psychologist and Consultant, milena.esherick@gmail.com, 0009-0000-0760-9183

Mohammed Al Owayyed

Delft University of Technology, m.alowayyed@tudelft.nl, https://orcid.org/0000-0002-9680-9204

Aušrinė Ratkutė

University of Twente, a.ratkute@student.utwente.nl

Gayane Sedrakyan

University of Twente, g.sedrakyan@utwente.nl, https://orcid.org/0000-0001-5045-5079

Mark de Reuver

Delft University of Technology, g.a.dereuver@tudelft.nl, https://orcid.org/0000-0002-6302-7185

Joao Fernando Ferreira Goncalves

Erasmus University Rotterdam, ferreiragoncalves@eshcc.eur.nl, https://orcid.org/0000-0002-8948-0455

Caroline A. Figueroa

Delft University of Technology, Stanford University, c.figueroa@tudelft.nl, https://orcid.org/0000-0003-0692-2244

* Corresponding author.

Youth increasingly turn to large language models (LLMs) for mental well-being support, yet current personalization in LLMs can overlook the heterogeneous lived experiences shaping their needs. We conducted a participatory study with youth, parents, and youth care workers (N=38), using co-created youth personas as scaffolds, to elicit community perspectives on how LLMs can facilitate more meaningful personalization to support youth mental well-being. Analysis identified three themes: person-centered contextualization responsive to momentary needs, explicit boundaries around scope and offline referral, and dialogic scaffolding for reflection and autonomy. We mapped these themes to persuasive design features for task suggestions, social facilitation, and system trustworthiness, and created corresponding dialogue extracts to guide LLM fine-tuning. Our findings demonstrate how lived experience can be operationalized to inform design features in LLMs, which can enhance the alignment of LLM-based interventions with the realities of youth and their communities, contributing to more effectively personalized digital well-being tools.

## 1 INTRODUCTION

There is an urgent need for more interventions to proactively support mental well-being of youth (ages 15-24) [1, 2]. Approximately three-quarters of mental health disorders emerging before the age of 25 and prevalence rates continue to rise globally [3]. Youth constitutes a formative developmental period marked by rapid physical, emotional, and social changes that both heighten vulnerability to health risks and offer meaningful opportunities for long-term prevention [4]. Digital platforms are an accessible medium for youth to seek mental well-being support, as they can offer immediacy and anonymity [5, 6]. Increasingly, these platforms incorporate custom chatbots (i.e., specialized automated agents) to deliver support [7]. In preventative mental well-being contexts, chatbots may assist in facilitating emotional check-ins, providing psychoeducation, and screening for early signs of distress [8]. However, despite growing implementation, many chatbots lack relevance and responsiveness to the complex and heterogeneous needs youth may face in their daily lives, limiting their potential for preventative support and digital mental health interventions (DMHIs) [2, 9–12].

One proposed solution is personalization, although personalization in digital health lacks a consistent definition. In practice, personalization in DMHIs is often understood as the use of persuasive techniques, such as tailored goal-setting, feedback, and reminders, to promote healthy behaviors in an individual [13–15]. These strategies are often integrated with evidence-based behavioral frameworks such as motivational interviewing, cognitive behavioral therapy, or the Behavior Change Wheel to ground interventions in individual contexts, such as for specific emotion states and time of day, to enhance and effectiveness [16–18]. However, current implementation of approaches can risk overlooking the complexity of lived experiences and the dynamic nature of daily circumstances, which is particularly salient for youth's evolving identities and contexts. In particular, a key limitation of many digital health tools, such as specialized chatbots, is their reliance on rule-based personalization, wherein personalization is implemented through fixed categories such as sociodemographic profiles, symptom thresholds, and/or predefined decision pathways [19–21]. While efficient, such approaches to personalization often lack personal relevance for youth and fail to capture the nuances of emerging mental health challenges.

Unlike rule-based digital health tools, Large Language Models (LLMs) generate adaptive responses based on the user's conversation, stemming from patterns observed during their training on large-scale datasets. This enables open-ended dialogue that can attempt to identify stressors and respond accordingly by drawing on conversational cues such as changes in language or sentiment [22–24], providing interactions that can potentially be perceived as more fluid and therefore personalized. However, given their novelty, crucial guidance is missing on how to deploy artificial intelligence (AI) tools such as LLMs that are relevant and grounded in the lived experiences of users, while mitigating potential harms. Such multiplicity of goals can be overlooked as current LLMs carry significant risk for youth mental health in particular, such as overreliance, alongside broader issues of hallucinations, algorithmic bias, marginalization, and lack of predictability [9, 25, 26].

Indeed, growing numbers of youth are turning to publicly available LLMs for well-being support, with recent surveys in the United States and European Union suggesting as many as three quarters of youth having used a messaging tool powered by LLMs, such as ChatGPT or personified chatbots, at least once [27, 28]. Such interactions often take place without oversight on commercially available platforms that are not explicitly designed or accounting for promotion of well-being. Documented cases reveal how vulnerable youth seek support from such platforms during severe mental health crises, raising concerns about the appropriateness and safety of such interactions [29, 30]. Critical questions remain unanswered about what LLM use for youth should entail, including the appropriate scope, depth, and boundaries of personalization, with expectations often differing sharply between youth, caregivers, and professionals.

For LLMs to meet their potential for enhanced personalization in well-being support while mitigating harms, there is increasing advocacy among scholars for co-creation of DMHIs, including LLMs, with youth and their communities of support (e.g., parents and youth care professionals) [31, 32]. A key challenge in leveraging LLMs for youth well-being is ensuring their responses reflect the values and lived realities of youth themselves rather than defaulting to assumptions embedded in training data or relying solely on professional guidelines. Achieving this requires community-grounded alignment, where youth and stakeholder perspectives can inform the fine-tuning of LLM toward relevance and appropriateness [33]. However, how to operationalize lived experience in LLMs and DMHI design remains unclear. Open questions include how to translate the ways youth experience distress, seek support, and define relevance into concrete dialogue features in personalized LLMs, as well as how safeguards can be co-designed to ensure that personalization remains supportive. These epistemic questions also highlight how lived experience can be considered a form of expertise based on firsthand knowledge, requiring systematic translation into technical design to advance LLM personalization that is relevant and responsive to users' personal needs. Further, to ensure implemented personalization mechanisms are effective in prevention and support in the long-term, there is a need to link participatory insights from stakeholders with evidence-based health strategies, including persuasive behavior change techniques [15].

To address these knowledge gaps, in this study we engage youth and community stakeholders to directly leverage their insights as lived experience experts to inform implementation of LLM personalization strategies in DMHIs for youth. Further, we focus specifically on co-creation with youth from marginalized socioeconomic backgrounds as a priority for population health and responsible LLM design. Our research questions (RQs) are:

RQ1: How can personalization in LLMs in preventative DMHIs for youth be grounded in the lived experiences of youth, parents, and youth care workers?

RQ2: How can these lived experience insights be operationalized to inform the design and alignment of personalization strategies in LLMs in preventative DMHIs for youth?

Importantly, we focused on preventative well-being or the prevention of acute mental health conditions [34]; therefore, clinical mental health, including diagnosis and treatment, was outside the scope of this study. In investigating these questions, we contribute to human-computer interaction (HCI) and digital health research by promoting a person-centered design approach which grounds individual and community lived experience in the co-design of evidence-based features and safeguards in LLMs in preventative DMHIs for youth. Taken together, these contributions advance HCI in three ways: conceptually, by reframing personalization as an ongoing dialogic and person-centered process grounded in lived experience rather than static tailoring; methodologically, by showing how youth-authored personas can act as person-centered scaffolds connecting personal, community, and health intervention perspectives with technical development; and practically, by offering concrete dialogue extracts and safeguards that can be incorporated into LLM fine-tuning pipelines, providing person-centered design guidance that is actionable for DMHI teams.

## 2 BACKGROUND

### 2.1 LLMs for Well-being Support

LLMs are increasingly integrated into DMHIs to support emotional well-being, foster self-awareness, and promote healthy behavior change [35–37]. Structured mental health tools leveraging LLMs are developed by multidisciplinary teams combining mental health expertise, LLM developers, and HCI design to provide sustained well-being support to diverse users. Existing examples include MindfulDiary, Wysa, and MindScape, which employ persuasive strategies such as behavioral prompts and goal-setting to maintain engagement, demonstrating improved well-being outcomes such as in emotional self-awareness and self-management [38]. Their support mechanisms can be systematically understood through persuasive system design (PSD), a comprehensive and evidence-based framework that categorizes features based on four types of well-being support (primary task, dialogue, social, and system credibility support) that are common in behavior change and information technology [15, 39]. For example, primary task support could mean step-by-step scaffolding to help a young person reach out to a friend, dialogue support might involve an LLM sending reflective questions, social support might involve presenting peer stories around well-being challenges, and system credibility support may reference real experts for health information.

Despite the existence of such evidence-based LLMs for structured DMHIs, they have not seen mainstream usage among youth. A 2025 review found that while existing DMHIs leveraging LLMs often employ dialogue support and primary task support features, many offer limited support for socialization or system credibility, such as mechanisms for promoting offline interactions or verifying health information [38]. This gap is particularly concerning for youth-facing interventions, where safeguards against misinformation, overreliance, or emotional harm, in addition to the need for referral to offline support, are critical. Furthermore, even digital health tools for youth that describe themselves as adaptive often reduce personalization to fixed categories such as socioeconomic background or quantitative symptom scores, rather than responding dynamically to the lived experience of end-users [19, 40]. As such, while existing DMHIs that use LLMs show promise, there is little guidance on how to design them for long-term personalization with diverse youth populations.

Conversely, growing numbers of youth are turning to publicly available, general purpose LLMs for well-being support. There is significant concern about how youth interact on these commercially available platforms, which are not explicitly designed for DMHIs. Further, although human feedback is incorporated in training and fine-tuning of LLMs on these platforms, this feedback typically comes from crowdsourced annotators, rather than in consultation community stakeholders with relevant lived experience expertise that can guide design [33, 41]. There have been multiple high-profile news cases involving parents suing companies that deploy LLMs, most notably OpenAI and Character.AI, following the suicides of their adolescent children, who had developed a parasocial bonds or sought inappropriate guidance during severe mental health crises on these platforms [29, 30]. Together, these events raise significant concerns on the risks of LLM tools that neglect community perspectives in their technical design and deployment; they also indicate the urgent need to improve LLMs developed for DMHIs as accessible yet evidence-based alternatives for youth.

### 2.2 Grounding LLM Personalization in Lived Experience Through Person-centered Design

Persuasive techniques improve engagement and health outcomes in personalized DMHIs. Yet, when mediated by adaptive intelligence such as LLMs, these same mechanisms raise ethical concerns and risks to youth well-being. AI systems that adapt dynamically to ongoing user input may unintentionally shape how youth interpret their experiences and decide when to seek external support, in ways misaligned with developmental needs or best practices in care, leading to algorithmic overdependence [42]. As such, guidance from professional psychological bodies and researchers emphasize the need to minimize manipulation, ensure developmental sensitivity, and ground design in real user needs in these systems [43, 44]. However, with rapid advances in generative AI such as LLMs, there is little practical guidance on how to scope and implement evidence-based persuasive features for health benefits while minimizing harm in leveraging this new technology for health promotion. Figueroa et al. [31] further argue that ethical guidelines for AI in youth mental health must be informed by the lived experiences of youth, especially from marginalized backgrounds, alongside insights from caregivers and professionals who support them daily. Without this grounding, tools risk being either overprotective, restricting usefulness, or insufficiently responsive to the realities youth face while perpetuating existing inequities in mental health access.

Grounding personalization of LLM-based support in DMHIs in lived experience means enabling youth to express themselves in their own words and receive supportive responses that reflect their concerns in their current context. To ensure relevance to youth in particular, such strategies must account for the dynamically shifting emotions, relationships, and goals that characterize their formative development [45]. This reframes persuasive mechanisms not as manipulation, but rather, as a supportive, ongoing personalization process that promotes well-being through strategies like reminders, motivational feedback, and reflective prompts which adapt based on specific user contexts over time [15]. Contextualizing persuasion through lived experience also responds to evidence that adaptive AI systems can foster harmful forms of personalization, with prior work on AI companion demonstrating how emotionally engaging, constantly available systems can blur human-computer boundaries and cultivate dependency in young people [25]. Grounding personalization in youths' lived realities and developmental contexts offers a path toward potentially mitigating these harms, as it reframes persuasive mechanisms as a means to promote reflection and autonomy rather than overreliance. Realizing this vision requires participatory and value-sensitive design processes that engage youth and their supporters in defining what genuine personalization and support looks like in practice and in the "real world". Specifically, participatory design emphasizes the direct involvement of communities in articulating how DMHI can enhance support and relevance, while value-sensitive design constitutes the systematic identification and integration of values of both direct users and the wider stakeholders who may be affected, such as parents and mental health professionals, into system design [46–49].

In sum, persuasive technology traditions help explain how LLMs in DMHIs can sustain both engagement and improved health outcomes [19,40]. To mitigate harms, participatory and value-sensitive design approaches can build on persuasive mechanisms by ensuring personalization is co-shaped with those who use and support it [31,45,48]. Together, these insights enable a person-centred approach to DMHI design, promoting "cooperative inquiry" with end-users as collaborators whose perspectives, values, goals, and contexts are central to intervention development [50, 51]. Beyond participatory design, person-centred design specifically emphasizes understanding the whole person, including their lived experiences and socioemotional realities, as foundational to aligning technologies with what genuinely supports them [52, 53]. By embedding lived experience throughout from the outset of technical design and implementation, a person-centred approach to design has potential to significantly enhance how LLMs in DMHIs are currently developed and deployed.

### 2.3 Eliciting Lived Experience Through Personas

Persona development, which began as a tool in user experience research in industry, have gained traction among health researchers as participatory tools for engaging stakeholders and grounding design in lived experience [54, 55]. In this study, we focus exclusively on personas that represent target population needs (i.e., of youth) to inform design; we do not refer to personas as implementation of personified conversational agents that embody characters [56]. We define personas as fictional character profiles grounded in empirical data such as surveys, interviews, and online forums to capture needs, behaviors, and lived contexts of end-users [55, 57, 58]. Personas extend traditions of person-centered representation in health intervention research by portraying the individual as a whole, situated within social, emotional, and environmental realities, rather than as a set of isolated attributes or symptoms. Personas can also serve a similar function to patient vignettes [53], by synthesizing relevant individual experiences to inform clinical intervention design and implementation; however, personas can be distinguished by their inclusion of factors beyond individual health, such as hobbies, values, daily schedule, and technological preferences, in the form of a holistic profile of a fictional person [59].

The process of developing personas through participatory engagement enables embedding user priorities directly into actionable design features, ensuring that system behaviors emerge from the lived contexts of intended users. Specifically, well-developed personas can inform contextualization, by capturing a holistic picture of a (fictional but realistic) person's daily life, challenges, relationships, and health status, and any other relevant aspects of their well-being and lived experiences [59]. Secondly, personas can enable operationalization, allowing stakeholders to envision concrete interactions between a person and a system and to translate complex qualitative insights into specific design features, all grounded in well-rounded user contexts [55]. For example, Bhattacharyya et al. [60] drew on patient and caregiver experiences with high-need, high-cost individuals managing multiple chronic conditions to create personas that helped clinicians, policymakers, and health system stakeholders align care priorities with the design of a personalized digital health assistant for these patients. In another study, individuals with perinatal anxiety and depression co-created personas to identify features for a DMHI tool, including transparent data practices and tailored resource navigation [61]. These cases illustrate how co-created user personas can function as triangulation tools, bringing intervention recipients, community stakeholders, and DMHI development teams together to refine system personalization requirements and safeguards.

Recent work has formatively explored applications of personas in LLM for health intervention contexts. For example, patient personas were employed as prompts to test one-time responses from GPT, highlighting the potential of user personas to enable adaptability of LLM outputs and improve upon communications written manually by healthcare professionals [62]. However, guidance on the integration of personas into LLM training pipelines, as well as how to create personas to inform design of LLMs for DMHIs, remain limited, especially for youth and preventative well-being. Given the nascence and lack of foundational guidance on developing LLMs for DMHIs, our current work focuses on leveraging personas as person-centered scaffolds for grounding iterative community alignment and qualitative inquiry, translating lived experience insights to inform the operationalization of LLM strategies to support youth mental health. This forms the conceptual and methodological groundwork for subsequent implementation research, where we aim to evaluate these co-created strategies and their integration into real-world mental health and community settings.

## 3 CURRENT STUDY

### 3.1 Overview of Methods

Guided by a person-centered participatory process, this study leveraged persona co-creation with youth and interviews with community stakeholders to thematically identify personalization strategies in LLMs for preventative DMHIs. Further, given persistent inequities in mental health outcomes faced by marginalized youth and their traditional exclusion from research [31], we actively partnered with community health organizations serving these populations, particularly youth from low income and socioeconomic backgrounds. We pre-registered the study protocol on Open Science Framework.[1] Based on persona co-creation and interviews, we developed a set of dialogue features and conversational strategies that reflected both stakeholder insights and evidence-based persuasive design principles to enable LLM personalization strategies to support youth mental well-being.

We leveraged persona co-creation with youth to scaffold broader community insight and their operationalization across four research stages. These stages include (1) scoping data-driven personas, (2) lived experience-based co-creation, (3) community-based interviews, and (4) translation to design strategies for LLMs in DMHIs. Each stage was implemented through collaboration with both the research (including technical) team and participatory stakeholders. Alongside Figure 1, Table 1 summarizes each stage, its rationale, implementation in the current study, and corresponding research question (RQ). We further detail each stage in the remaining section.

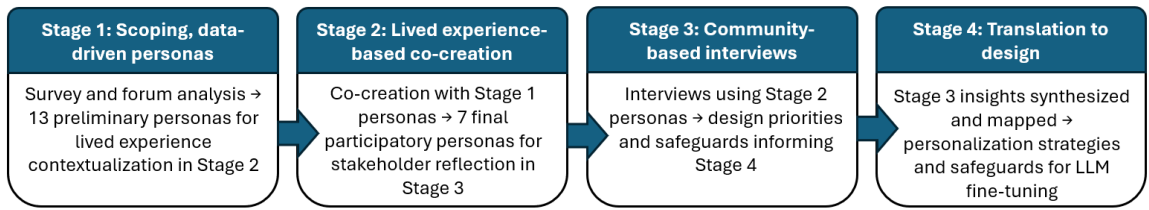

Figure 1: Our four-stage procedure to leveraging lived experiences of community stakeholders in developing LLM personalization strategies and safeguards.

---

[1] https://osf.io/khfxu/

Table 1: Overview of Stages, Goals, Methods, and Outputs in Current Study

| Stage | Goal | Method | Key Output | RQ |
|---|---|---|---|---|
| 1. Scoping, data-driven personas | Ground co-design in well-being considerations from existing self-reported experiences | Clustering of global mental health coping strategies survey; thematic analysis of youth forum posts | 13 preliminary personas (10 survey-derived, 3 forum-derived) for participatory refinement in Stage 2 | RQ1 |
| 2. Lived experience-based co-creation | Refine, contextualize, and extend scoping personas through youth input based on their first-hand lived experiences | 3 youth workshops (N=24) involving the critique of Stage 1 personas and refinement, with youth creating their own "participatory personas" | 7 "participatory personas" reflecting youth well-being needs and lived experiences as basis for person-centered, community-based co-creation in Stage 3 | RQ1 |
| 3. Community-based interviews | Elicit person-centered design insights and LLM personalization opportunities from community members | Think-aloud interviews (N=14) with youth, parents, and youth care workers using Stage 2 participatory personas as person-centered discussion stimuli and with questions from adapted Persona Perception Scale (PPS) | Person-centered considerations for LLMs for youth DMHIs, directly informing exemplary LLM personalization strategies and guidelines in Stage 4 | RQ1/ RQ2 |
| 4. Translation to design | Translate lived experience into LLM personalization strategies | Thematic coding of Stage 3 findings and persuasive features mapping; example dialogue scaffolds | LLM-ready dialogue extracts and design safeguards to embed into training and fine-tuning pipelines | RQ2 |

*3.1.1 Stage 1: Scoping, data-driven personas*

To ground subsequent co-design stages in existing evidence as a starting point, we first developed preliminary "scoping" personas from existing online sources. These personas served as scaffolds to prompt reflection, critique, and personalization in later workshops. We used Survey2Persona [63], a machine learning platform created by HCI researchers, to cluster a subset (N=6,272) of the Wellcome Global Monitor: Mental Health survey[2] focused on youth aged 13–25 in Europe or top immigrant countries. Personas were quantitatively clustered by survey reports related to well-being coping strategies, such as talking to friends, making lifestyle changes, engaging in spiritual practices, or seeking clinical support. This process generated 10 unique personas, each reflecting a distinct coping profile. We used default parameters and did not experiment with clustering variations, as the goal was to create useful starting points for later co-design stages, rather than create finalized persona profiles for direct LLM system implementation.

To supplement the quantitative personas, we thematically analyzed 100 public posts from Kindertelefoon[3], a Dutch youth helpline forum. Posts covered common well-being concerns such as friendship conflict, school stress, and family issues. Based on this analysis, youth researchers created three additional personas to capture common youth contexts to supplement the survey data on coping strategies. In total, 13 personas were created: 10 from survey clustering and three from forum themes (see Supplement for examples). These personas were introduced in the next stage as co-design scaffolding tools for youth to critique, adapt, and expand based on their own lived experiences.

---

[2] https://wellcome.org/reports/wellcome-global-monitor-mental-health/2020
[3] https://www.kindertelefoon.nl/

*3.1.2 Stage 2: Lived experience-based co-creation*

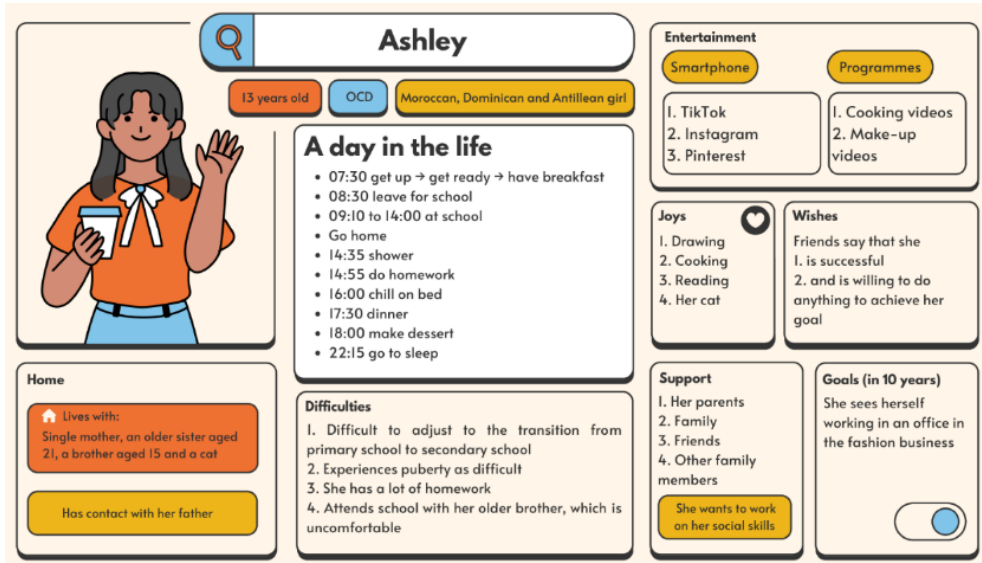

Figure 3: Example of participatory persona created by youth during co-creation workshop, converted from paper to digital verbatim. Youth were provided with a blank structured template to guide, but not restrict, their creation of a representative persona to inform LLM personalization strategies.

The second stage focused on refining the 13 preliminary personas from stage 1 through participatory co-design workshops with youth to ensure relevance and grounding in their lived experiences. All workshops were held at Talenthub Op Zuid[4], an organization in the Netherlands that supports youth at risk of dropping out of vocational training, including youth from lower socioeconomic backgrounds; this enabled a diverse participant sample that is traditionally overlooked in digital health design. We conducted three 90-minute workshops at Talenthub Op Zuid, each with a different group of participants (total N=24, MAge=17.6, SD=1.2, see Supplement for additional details).

In the first workshop, participants reviewed the prior 13 personas from Stage 1 and critiqued them for gaps in relevance. The scoping personas generated from survey and forum data gave youth stakeholders a concrete starting point for consulting as experts by experience in initial co-design activities. They challenged the realism of the scoping personas. Using fill-in-the-blank templates to guide but not restrict their persona creation (created by a youth member of the research team with design training, see Supplement), youth added contextual details to the project personas, such as daily routines, stressors, and digital habits, and brainstormed plausible backstories involving bullying, school difficulties, or parental conflict. The second workshop engaged a new participant group who expanded on previous outputs and addressed additional questions on living environment and emotional support needs, as this was suggested as relevant by youth from the prior workshop. Participants revised or created new personas based on their own or peers' experiences. In the third workshop, a new group of participants again reviewed prior co-creation and outputs and further refined the personas.

---

[4] https://talenthubopzuid.nl/

This iterative process across three complementary workshops resulted in seven "participatory" personas reflecting diverse youth lived experiences across personal, social, daily life, and digital domains and capturing diverse lived experiences across backgrounds, relationships, aspirations, and health concerns. These participatory personas were used in the subsequent stages to inform stakeholder interviews for identifying personalization features in line with youth's lived experiences. Figure 3 provides one example of a participatory persona ("Ashley").

*3.1.3 Stage 3: Community-based interviews*

To directly answer how lived experience can inform LLM personalization strategies (RQ1) and enable their translation to system design (RQ2), we next conducted semi-structured, think-aloud interviews (30-45 minutes) with 14 additional community stakeholders: new youth participants (not involved in prior workshops), youth care workers (adults who support youth well-being in community settings), and parents (see Supplement for additional details). During these interviews, the youth-created participatory personas from Stage 2 served as scaffolds to anchor interviewees' reflections and recommendations for LLM personalization.

Each interviewee reviewed the final set of seven participatory personas and responded to questions adapted from the constructs in Persona Perception Scale (PPS) [64]. The PPS is a validated set of questions that provide a comprehensive assessment of dimensions of representativeness and relevance in personas for design research, such as credibility, consistency, and completeness [64]. We adapted PPS questions for our current use case. Specifically, our interview questions explored how the personas co-created with youth might usefully inform LLM behaviors, personalization, and safeguards. Interviewees reflected on what was captured well, what was missing, and how an LLM could or should not act on information in the personas. Sample questions included: "What would you want an LLM to know about this persona or other youth that isn't included?", "What could go wrong in this interaction with real youth?", and "What should the LLM ask this persona or other youth to personalize support safely?"

*3.1.4 Stage 4: Translation to design*

To answer RQ1 and RQ2, we conducted a two-phase analysis of the stakeholder interviews from Stage 3. First, inductive thematic analysis was used to identify key personalization strategies for well-being tools that leverage LLMs based on the lived experiences of stakeholders (RQ1). We followed Braun and Clarke's six-phase reflexive thematic analysis method [65]. Three researchers independently coded all interview transcripts to extract "in vivo" codes (direct quotes verbatim) representative of stakeholders' lived experience perspectives. Through iterative review with the rest of the research team, in vivo codes were grouped into broader categories and refined into overarching themes. Two additional researchers reviewed the transcripts and validated the final themes. Coding and synthesis were supported by ATLAS.ti and Excel.

Once thematic saturation was reached, the themes were deductively mapped onto relevant PSD features (defined in the Background section) and translated into exemplary dialogue extracts, operationalizing lived experience insights into concrete evidence-based personalization strategies and safeguards in LLM responses (RQ2). A researcher specializing in PSD compared each inductive theme with established PSD categories to map alignment between stakeholder insights and existing behavior change communication strategies. Stakeholder themes that extended beyond existing PSD categories (for example, concerns about privacy) were unmapped.

All themes (both PSD-mapped and non-mapped) were then translated into exemplary dialogue extracts or system guardrails. For PSD-mapped themes, exemplary dialogues were constructed by aligning stakeholders' verbatim quotes (the in vivo codes from inductive analysis) with established formulations of the corresponding PSD feature (e.g., reminders) from prior literature. For themes that could not be mapped to PSD, exemplary dialogue extracts were constructed directly from stakeholders' verbatim quotes. This process ensured that the resulting LLM personalization strategies were both grounded in lived experience and consistent with evidence-based design principles, while also capturing new considerations and values from stakeholders. Together, the dialogue outputs resulting from our stakeholder themes provide concrete guidance for personalizing well-being tools that leverage LLMs. In future work, we envision exploring how the dialogue extracts and safeguards identified in this study can be used as material for LLM fine-tuning, which was out of the current scope of the study.

### 3.2 Positionality and Reflexivity

Our team included researchers with expertise in psychiatry, clinical psychology, behavioral science, youth engagement, digital health, public health, computational social science, design, and information systems. We also brought diverse sociodemographic backgrounds and personal experiences with health systems, mental health, and marginalization, which shaped how we approached questions of well-being and personalization. This mix of lived experience and professional expertise informed both our study design and goals to capture youth and community perspectives and ground their insights in design decision-making. To capture their perspectives as authentically and respectfully through open dialogue, all co-creation workshops and interviews were facilitated by younger members of the research team (under 25), who were also involved in validation in findings. We also piloted co-creation workshops with youth care workers and designer professionals outside the study to refine activities and ensure they were engaging and appropriate for youth participants. Wherever possible, the current findings aim to (anonymously) report stakeholders' insights verbatim to ensure representativeness.

## 4 FINDINGS

### 4.1 RQ1: Results of Thematic Analysis of Lived Experience Insights

To address our RQ1, our analysis first focused on identifying insights into personalization in LLMs for DMHIs from the perspective of community stakeholders' lived experiences. Each interviewee viewed all seven participatory personas in full to guide the interview discussion. The personas' realism resonated with some ("Ashley reminds me of my sister", "I know many Jims and Davids"). However, all interviewees consistently stressed that details in the personas, while useful as starting points, were insufficient for personalization and required the LLM to probe further what those attributes meant in practice; for example, identifying a persona simply as a 19-year-old woman, as Christian, or as having OCD. They also highlighted omitted details in certain personas such as school level, professional track, daily routines, social media use, or hobbies, noting that such information would critically shape personalization of DMHI support to be grounded in lived experience. Stakeholders further emphasized that LLMs would need to probe beneath surface statements to uncover underlying causes, such as whether financial stress stemmed from gambling or family pressures, or whether friendship difficulties reflected low self-esteem or social skill challenges. These reflections underscored how the participatory personas scaffolded missing details and understanding their significance for lived experience-informed personalization, providing the basis for our thematic analysis across stakeholder interviews in Stage 4 to answer RQ1. Full quotes are presented in Tables 2-4 (labeled by stakeholder group and a numerical identifier to differentiate individual interviewees, e.g., Youth 3, Parent 2).

The reflective engagements with personas enabled thematic analysis of stakeholders' lived experience perspectives related to LLM personalization. From this analysis, three overarching design priorities for personalized DMHIs for youth that leverage LLMs emerged. First, personalization prompts should continuously probe for underlying meaning and contextual causes rather than rely on inference. Next, safety boundaries around LLM content scope, role, and authority should be made explicit across interactions. Finally, stakeholders outlined several dialogic scaffolding strategies to guide LLM questioning in ways that provide relevant yet safe personalization. We synthesized these youth, parent, and youth care worker perspectives into these three interrelated themes of lived experience personalization to answer RQ1: *Contextual Understanding as the Core of Personalization, Safety Boundaries in Personalization*, and *Dialogic Scaffolding for Reflection and Autonomy.* For each theme, we also denote subthemes and LLM design considerations capturing distinct needs and priorities (Tables 2, 3, and 4 respectively).

*4.1.1 Theme 1: Contextual Understanding as the Core of Personalization*

Table 2. Subthemes for Contextual Understanding in LLM Personalization

| Subtheme | Definition | Illustrative quotes |
|---|---|---|
| Beyond demographic labels | Move beyond demographics or diagnoses and probe for lived experience. | "Personalization is about understanding what makes a person who they are." – Youth 3<br><br>"The LLM shouldn't just think 'this is a nineteen-year-old woman,' but recognize her particular goals in life." – Youth 5<br><br>"This persona says she's Christian, but many people identify as Christian without really believing." – Parent 3<br><br>"OCD manifests differently in everyone. Does this mean she has learning disabilities? Trouble concentrating? You don't know what it means for Ashley [the persona]." – Youth care worker |
| Educational and literacy context | Account for educational level and literacy constraints to ensure accessibility. | "What's this persona's [Dutch educational] level — MAVO, HAVO, or VWO? That frames the issue and how to respond." – Youth care worker 4<br><br>"For an LLM to help, I think we need to know more about this persona, about their background, their education." – Parent 3 |
| Nuanced causes of common challenges | Probe for the diverse underlying causes of problems. | "You're tired, not having fun, maybe being bullied — there are many reasons why you don't want to go to school." – Youth 3<br><br>"What does she need to make friends? Is it a lack of social skills, fear of starting contact, or is she just very deep and doesn't want to be superficial?" – Youth care worker 2<br><br>"If they say their financial situation is difficult, the LLM needs to find out why. Are you spending a lot of money? Then the LLM understands whether it's being spent on gambling or alcohol or shopping addiction or whatever." – Parent 2 |
| Relational dynamics over roles | Explore relationships by focusing on effects. | "The LLM should ask about people like parents and friends, but not just who they are which doesn't really matter, but what they are like. If her mother is always controlling her, putting a lot of pressure on her for school, then that affects Ashley the persona." – Youth 4<br><br>"I think she has very low self-esteem, which makes it difficult for her to make friends. I wouldn't say it should be about the latter. So the LLM should figure out if she has low self-esteem." – Youth care worker 4 |

| Subtheme | Definition | Illustrative quotes |
|---|---|---|
| Everyday routines and (digital) practices | Draw on habits, hobbies, and digital practices to infer motivations and tailor suggestions. | "Her hobbies already reveal 50% of who she is." – Parent 4<br><br>"It would be nice to know his hobbies or what makes him happy. That gives the LLM a hint about where to start. If he enjoys baking, maybe it could suggest a cooking class to meet friends." – Youth 1<br><br>"I don't see social media or personal relaxation being factored into that persona." – Youth care worker 4<br><br>"What you watch on social media isn't who you are, but it does show what you like." – Youth 4 |
| Dynamic adaptation | Adapt to fast-changing youth contexts and update guidance accordingly. | "She was close to her grandparents, but not anymore. That changes her support network and the advice she needs." – Parent 3<br><br>"Teenagers can be super happy one moment and very sad an hour later." – Youth care worker 5<br><br>"If someone felt bad a month ago but is fine now, the LLM might still act as if they're not okay." – Youth 4<br><br>"It's very black and white: these are challenges, these are goals. But a teenager's life isn't always that simple — it's dynamic." – Youth care worker 3 |

As seen in Table 2, stakeholders described personalization as developing *Contextual Understanding*, which requires attention to the everyday circumstances that shape how youth experience challenges and support, by going beyond demographic labels to probe individual goals, identities, and the unique ways conditions manifest. Participants rejected surface-level tailoring in DMHIs based on static categories such as age, religion, or diagnosis, emphasizing instead a more educational and literacy-aware approach that considers differences in background and comprehension as central to accessibility. Everyday routines and digital practices such as hobbies, online behaviors, and social media use were also identified as critical inputs, providing insight into motivations and entry points for personalized support.

Stakeholders also highlighted the need to recognize nuanced causes of common challenges, noting that similar outward behaviors, such as academic stress or social withdrawal, could arise from different sources, ranging from low self-esteem to fear of superficiality or family pressure. They argued that systems should uncover these distinctions through dialogic questioning (see Theme 3, Table 4 findings on suggestions for dialogic scaffolding features) rather than assume common root causes. Social connections were described as central to well-being, but participants stressed that it was the relational dynamics over roles that mattered most: not just simply whether a parent, peer, or sibling was present, but whether those relationships were experienced as supportive, controlling, or esteem-building. Finally, participants emphasized the importance of dynamic adaptation. Adolescents' contexts were seen as fluid, with moods, motivations, and support networks shifting rapidly in ways that altered needs. Without mechanisms for re-personalization, the DMHI risked responding to outdated circumstances, continuing to address distress even after recovery. Stakeholders emphasized the need for DMHIs to update assumptions regularly via an LLM and adjust support in line with a young person's evolving situation and lived experience.

*4.1.2 Theme 2: Safety Boundaries in Personalization*

Table 3: Subthemes For Safety Boundaries in LLM Personalization

| Subtheme | Definition | Illustrative quotes |
|---|---|---|
| Privacy and disclosure | Recognize that youth may limit or misrepresent input due to fears of surveillance or data misuse. | "You don't know who's behind this [LLM]… it could end up with your parents." – Youth 3<br><br>"Some young people downplay or misrepresent interests, like hobbies, because they're cautious about trusting AI." – Youth care worker 4 |
| Content and behavioral boundaries | Avoid overconfidence, avoid sensitive topics, and mitigate risks of overreliance. | "When I ask AI if there will be a World War 4, it answers. I find that scary." – Youth care worker 5<br><br>"A thirteen-year-old girl doesn't know what's acceptable and what's not, so the LLM shouldn't present certain things." – Youth care worker 4<br><br>"It's there to listen to you and maybe offer advice, but we don't want young people becoming addicted to talking to such a tool." – Youth 1<br><br>"When AI guides them to recognize their problems, they can acknowledge it. But if it starts giving specialist answers, like a doctor would, you could send someone in the wrong direction." – Parent 2 |
| Referral to offline resources | Frequently redirect to offline resources, escalating appropriately when human connection is essential. | "If they're struggling with something, you should be able to give them a hug — an LLM can't do that. Let's keep it human, keep it personal." – Youth care worker 3<br><br>"I would include a social map in the AI tool if the situation is serious. Then the tool can filter and refer someone to an organization or foundation, so they engage at a more personal level than just with a chatbot." – Youth 1 |
| Use of peer narratives | Leverage authentic cases from similar situations faced by peers to ground advice in lived experience. | "I think you also need people who already had problems, and the LLM knows how they solved it, so it can recognize that. For example, I often see that people have few friends. Someone with few friends can tell how they learned to make friends." – Youth 2 |
| Transparency and role clarity | Make clear its non-human role, state limits, and explain how advice quality depends on user input. | "Without enough context, advice comes out vague and full of assumptions, which could make the system make things up. I think you should see AI as AI. It either has to be very clear or very general. Then AI can ask you to be more specific." – Youth care worker 3<br><br>"The more you give the LLM, the more concrete the advice you get back." – Youth 1 / Parent 1 |
| Neutral tone | Avoid mimicry of compassionate human responses, stay neutral. | "I find an LLM useful because they're less biased. When you go to a therapist, you pay them. Of course, they say you're doing well, even though that's not the case at all. That's too sweet. A neutral critical tone I find useful." – Youth 2 |

Stakeholders also stressed that personalization in youth DMHIs must operate within clear limits to avoid harm as *Safety Boundaries* (Table 3). Privacy and disclosure emerged as a central concern, with youth focusing less on abstract policies and more on whether they could trust the system to keep confidentiality over what they shared. Participants also underscored the importance of transparency and role clarity, noting that systems must make their limits explicit and communicate clearly what data is stored, who controls it, and how responses depend on user input. Participants called for strict content and behavioral boundaries, highlighting the risks of overconfident or speculative responses on sensitive issues and cautioning against forms of interaction that could foster overreliance. When topics exceeded what an AI could responsibly handle, they stressed the need for referral to offline resources, directing young people to human services and support. At the same time, some identified the value of drawing on peer narratives, suggesting that authentic lived examples could ground advice and make guidance more relatable. Finally, there was broad agreement that LLMs should adopt a neutral tone, maintaining a critical rather than overly compassionate style. This neutrality was seen as essential to safeguard autonomy, prompting reflection and clarification in DMHIs rather than steering young people toward particular choices.

*4.1.3 Theme 3: Dialogic Scaffolding for Reflection and Autonomy*

Table 4: Subthemes For Dialogic Scaffolding in LLM Personalization

| Subtheme | Definition | Illustrative quotes |
| --- | --- | --- |
| Intent clarification | Continuously ask end-user to define scope. | "I don't know what she [this persona] does. The LLM should help clarify what exactly she wants to do." – Youth 5 |
| | | "If someone asks me something like what I do in my life, I say kickboxing and school. I do that, but that's not what I only actually do in a day. So the LLM should always ask 'what else' right away." – Youth 2 |
| Interpretive questioning | Use questions to explore personal meaning and reflection. | "What matters is not the word 'difficult,' but how the persona saying this finds it difficult." – Youth 1 |
| | | "The LLM can use an hourglass model, where you start by asking all sorts of questions until you get to the core. This will help you increase young people's reflective capacity." – Youth care worker 2 |
| Narrative and goal exploration | Use stepwise questioning to link short-term challenges to deeper aspirations. | "This persona says she likes her eyes, but would like to change something about her appearance. How badly does she want that? Is that something that really stands in her way? Or is it something she actually would like, but which she won't do when she really thinks about it? What is the weight like? Everything has a certain weight. How is AI going to recognize that weight without that information?" – Youth care worker 2 |
| | | "So that the chatbot is trained well in helping someone, it should determine who, what, where, when, why." – Youth 2 |
| Autonomy support | Facilitate exploration while leaving final choices to youth. | "I wouldn't go to an AI to make a choice about something." – Youth 5 |
| | | "It's important to show young people it's okay to make mistakes or to be patient. You don't want to radicalize them into success." – Youth care worker 3 |

Stakeholders described personalization as a process of structured dialogue that empowers youth think through their situations, rather than a DMHI system that delivers ready-made answers through an LLM. Stakeholders emphasized four strategies as central to *Dialogic Scaffolding* in personalization (Table 4). Intent clarification was viewed as essential for aligning the scope, tone, and depth of exchanges with a young person's stage of development and readiness. Once this foundation was established, interpretive questioning was valued for helping adolescents articulate the personal meaning of abstract concepts, encouraging them to define terms in their own way rather than accept generic interpretations. Building on this, narrative and goal exploration was identified as a key potential scaffold, consisting of stepwise prompts that connect immediate concerns to broader values and aspirations, situating short-term difficulties within longer-term trajectories. Participants further highlighted the importance of autonomy support, stressing that personalization from LLMs is most effective when it guides reflection and recognition of patterns while leaving choices to the young person, rather than offering prescriptive or definitive answers.

**4.2 RQ2: Translating Themes to LLM Design**

To address RQ2, we mapped these subthemes based on stakeholders' insight onto persuasive system design (PSD) categories and features (see Background section) and translated them into exemplary dialogue extracts (summarized in full in Table 5). This enabled demonstration of how lived experience can be operationalized into evidence-based behavior change strategies and design requirements in LLMs used for DMHIs.

Several subthemes aligned directly with existing PSD categories and features. For example, the subthemes *beyond demographic labels* and *educational and literacy context* were mapped to the PSD Primary Task Support (PTS) feature of tailoring, which was operationalized in dialogue extracts that ask youth what the DMHI's LLM system should know to be most helpful or allow them to select preferred literacy levels for responses. *Narrative and goal exploration* aligned with PTS features of tunneling and reduction, where large tasks such as texting friends can be broken into smaller steps to make social goals more manageable. Meanwhile, operationalization of the subtheme, d*ynamic adaptation* drew on both PTS and Dialogue Support (DS), with dialogue extracts presenting reminders and suggestions that adapted in real time to user input, such as prompts to update a mood log.

Relational subthemes were primarily mapped to Social Support (SS) features. For example, the subthemes *relational dynamics* and *peer narratives* informed dialogue extracts that highlight peer experiences by sharing anonymized youth forum stories. These strategies drew on social learning and social facilitation features to situate individual challenges within wider social and offline contexts. Similarly, the subtheme *referral to offline resources* combined both SCS with DS features, with corresponding dialogue extracts that direct youth to trusted helplines or local services when needed.

Most personalization safeguards recommended by community stakeholders were mapped to System Credibility Support (SCS) features. For example, *privacy and disclosure* informed dialogue extracts that explained data handling and reinforced user control. The subtheme of *content and behavioral boundaries* aligned with trustworthiness and real-world feel features, reflected in extracts that clarified the limits of system roles (e.g. "I am not a human so there are limits to how much I can help, but a real person definitely can"). Similarly, the subtheme *Transparency and role clarity* was operationalized in dialogue extracts that clarified system limitations while also explaining why information was being requested.

At the same time, several subthemes extended beyond PSD. These were not easily captured as persuasive features and were instead developed directly into dialogic scaffolding strategies. For example, the subtheme *nuanced causes of common challenges* reflected the need for systems to ask open questions that help youth explore why difficulties are happening rather than offering solutions. *Interpretive questioning* encouraged youth to generate their own explanations, supporting meaning-making instead of persuasion. *Intent clarification* focused on beginning each exchange by asking the youth what they wanted to discuss, ensuring the conversation followed their agenda. *Autonomy support* was expressed through scaffolds that consistently left choices open, so that suggestions were always paired with user feedback and opportunities to explore alternatives. Finally, *neutral tone* was identified as important for safeguarding, with stakeholders warning against exaggerated praise or simulated empathy that could feel inauthentic or lead to overreliance.

Together, the dialogue extracts and dialogic scaffolding strategies presented in Table 5 can contribute directly to the supervised fine-tuning phase of LLMs for DMHIs. Unlike generic fine-tuning data or crowdsourced annotations, these current study examples are grounded in the lived experiences of youth, youth care providers, and parents. They provide a resource for embedding community perspectives directly into DMHI design, moving LLMs from providing generic support or safety toward personalization that adapts to the specific contexts of youth well-being.

Table 5: From Stakeholder Insight to Dialogue Extracts and Strategies for LLM Personalization in DMHIs

| Subtheme from prior section | Relevant PSD features | LLM dialogue extract or strategy (if no relevant PSD feature) |
|---|---|---|
| Beyond demographic labels | PTS: Tailoring | "What are the most important things I should know about you in order to help you?" |
| Educational and literacy context | PTS: Tailoring | [first time a conversation occurs, present three responses in varying literacy level to guide all future exchanges] "Rate which of these three responses you prefer." |
| Nuanced causes of common challenges | N/A | "Why do you think you are avoiding school?" |
| Relational dynamics over roles | SS: Social learning | "You mentioned your friend has been reaching out to hang out recently—would you like tips on how to invite them to join you?" |
| Everyday routines and (digital) practices | DS: Reminder<br>DS: Suggestion | "You've mentioned previously that you enjoy art in your free time — would you like a suggestion to use it as a quick wind-down after school? |
| Dynamic adaptation | PTS: Tailoring<br>DS: Suggestion | "Have you filled out your mood chart in the app today? This can help determine how to best help you at the moment." |
| Privacy and disclosure | SCS: Trustworthiness | "As an LLM I can also access stories from people who faced similar situations. Would you like me to share their story with you without sharing identifying information?"<br><br>"Here is how your data is handled and stored. You are in control of what you share. Are you okay to continue?" |
| Content and behavioral boundaries | SCS: Trustworthiness, Real-world feel | "I am not a human so there are limits to how much I can help you, but a real person definitely can." |
| Referral to offline resources | SCS: Trustworthiness, Expertise, Real-world feel<br>DS: Suggestion | "In the Netherlands, Kindertelefoon is a free and anonymous helpline to support young people. You can discuss anything on your mind. It can be reached by calling 0800-0432 or through their website for online chat. They are available every day from 11:00 to 21:00."<br><br>"I suggest you talk to a real person right away. Your mood is lower than usual (usually you rate a 7 out of 10) and it's important you get help in case it's urgent. Would you also like a list of nearby community centers you can drop-in today for support? " |

| Subtheme from prior section | Relevant PSD features | LLM dialogue extract or strategy (if no relevant PSD feature) |
| --- | --- | --- |
| Use of peer narratives | SS: Social learning, Social facilitation SCS: Real-world feel | "You've mentioned previously that you enjoy art. I found a relevant story based on my training from forum posts on Kindertelefoon. So a poster once shared that after his breakup, he felt lost and isolated. So he joined a local art class which helped him process his feelings. He said that while creating art, he reflected about what had happened and how he felt about it. He was surprised to find that he felt better when he was done with the paintings." |
| Transparency and role clarity | SCS: Real-world feel | "As an LLM I am not a human so I have not experienced this. But I know a helpline where you can talk to someone anonymously." |
| | | "We have better conversations when I know how you are feeling, because then I have more information to help you at a specific moment. That's why I ask regularly." |
| Neutral tone | N/A | Strategy: Minimizing amount of praise and empathetic language in training data. |
| Intent clarification | PTS: Tunneling | Strategy: Every exchange starts with the LLM first asking what the user wants support with. |
| Interpretive questioning | N/A | Strategy: Asking "why" frequently during exchanges, e.g. "Why do you think you are avoiding school?" |
| Narrative and goal exploration | PTS: Tunneling, Reduction | "This can be broken down step by step. First, decide on the tone you want to set — casual, warm, or direct. Then, draft one or two short messages that match the tone you want. Finally, send one when you're ready, so it feels less overwhelming and you get it done." |
| Autonomy support | N/A | Strategy: each LLM turn should end with a suggestion, followed by asking the user what they think of it and if they would like an alternative. |

## 5 DISCUSSION

Personalization in digital health is often delivered through rule-based categorizations, yet such approaches limit both relevance and responsiveness [19–21]. The rapid emergence of LLMs marks a critical seachange, as these systems can generate adaptive responses from conversational patterns, enabling open-ended dialogue with end-users [22–24]. Yet their novelty raises important questions about how LLMs personalize digital mental health interventions (DMHIs) through relevance to diverse lived experiences while mitigating risks, such as misinformation and overreliance, particularly among youth [25]. Many existing commercial LLMs overlook developmental needs and seldom incorporate community perspectives [29, 30]. In parallel, digital well-being tools that leverage AI such as LLMs increasingly employ persuasive design; however, guidance remains limited on how to ensure these strategies support well-being rather than risk manipulation [31, 43]. These multiple gaps are particularly concerning given the growing reliance of youth on tools that leverage LLMs for personal well-being support [27], prompting calls for participatory and value-sensitive design with communities [31, 32].

In response, in this study, we asked two related questions: first, how personalization in LLM-based youth well-being tools can be grounded in the lived experiences of youth, parents, and youth care workers (RQ1); and second, how these participatory insights can be operationalized into concrete design strategies and alignment features (RQ2). Our study demonstrates how DMHIs leveraging LLMs can begin to address current challenges in responsible AI by embedding person-centered co-creation in their design, and we propose a person-centered model of designing DMHI personalization in collaboration with end-users and community members (Figure 2). Specifically, our methods and findings indicate how personalization of youth DMHIs can be grounded in lived experience and holistic personal experiences, by systematically operationalizing community stakeholders' insights from co-creation into evidence-based health behavior change and strategies and dialogic scaffolds, in turn providing actionable resources for implementing LLM alignment and fine-tuning for DMHIs. In the following section, we consider theoretical and practical implications of our work for different stakeholders, as well as its limitations and directions for future research.

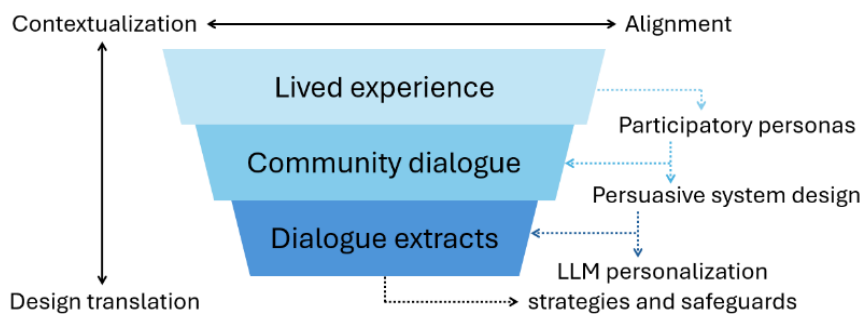

Figure 2. Overview of the current study's person-centred approach for translating lived experience into system design. Lived experience insights were facilitated through participatory personas co-created with youth and community dialogue, then mapped to persuasive system design to generate dialogue extracts to inform LLM personalization. The figure illustrates how this process links contextual grounding with system alignment, while also moving from lived experience toward design translation.

## 5.1 Theoretical Implications: Designing for Person-centered Personalization

Our findings extend HCI theories of personalization by reframing it as a dialogic and person-centered process that develops through ongoing interaction and adaptation. Prior work in digital health has often conceptualized personalization as a means of enhancing engagement, using persuasive techniques for behavior change or adapting responses based on predefined behavioral variables [13, 39, 40, 62]. Building on situated and value-sensitive perspectives in HCI [46, 48], our current work instead proposes that LLM-driven personalization emerges in interactional dialogue with end-users, rather than through pre-specified adaptation rules [40, 66], to foster personal meaning. Specifically, participants across our co-creation sessions valued the role of an LLM to facilitate narrative exploration through stepwise questioning, which can allow youth to meaningfully articulate immediate concerns and connect them to broader aspirations related to their mental well-being in daily life. Likewise, stakeholder insights on the importance of everyday routines, social experiences, and fluctuating emotional states demonstrated how LLM support can be contextualized on a personally meaningful level, such as on days with low mood reports, by prompting youth to reflect on whether they had engaged in aforementioned favorite hobbies, and adjusting subsequent activity suggestions or resources accordingly. In this way, digital health personalization can move beyond static tailoring to become an interactional process grounded in lived experience and person-centered dialogue, encouraging individuals to explicitly relate the content of their interactions with an LLM to offline context and personal meaning.

Crucially, these insights position LLM personalization in DMHIs not as a system for automated support, but as a mechanism for dialogical scaffolding of self-development. Participants valued space for self-reflection and discovery over prescriptive advice, envisioning LLM-based DMHIs that, even when retaining memory of personal details of the user, primarily ask questions and maintain a neutral, inquisitive tone rather than offer affirmation. Such an approach supports individuals by enabling them to support themselves, including directing to relevant peer narratives, alongside offline or community resources. Such mechanisms for personalization are particularly salient when LLMs are used by youth, who are still developing autonomy, self-management, social and decision-making skills [4]. At their developmental stage, youth may be especially susceptible to overreliance on external or problematic algorithmic guidance [25, 43], highlighting the need for LLM interactions that cultivate reflective capacity, self-efficacy, and social belonging as a foundation for long-term well-being, instead of only providing information in response to queries. For youth from marginalized communities in particular, who face systemic barriers to health access and societal participation [68], such dialogical personalization may especially help foster agency and belonging. Specifically, dialogical support, combined with retrieval of relevant peer narratives and referral to community resources can help marginalized youth better explore and navigate solutions for their mental well-being.

Importantly, any referral to peer or community resources by an LLM must remain grounded in the individual's lived context, including their socioeconomic capacity, as indicated by our community experts and prior literature on digital health literacy [69]. Rather than offering generic or prescriptive materials, our findings indicate significant potential for enhancing the precision of LLM resource retrieval capabilities to adapt from ongoing dialogue and reflect a person's uniquely articulated needs and circumstances, as seen in recent information retrieval research on retrieval-augmented generation (RAG) [70, 71]. In this way, LLM-based DMHI personalization strategies do not begin with predefined categories of support but from within dynamic user interactions. Taken together, the current study insights reposition personalization within HCI and digital health as a process of co-inquiry linking digital interaction with reflection and real-world action to enhance self-management and well-being.

Our work also advances methodological conceptualizations of HCI design artefacts, demonstrating how personas authored directly by end-users — in this current work, youth from marginalized backgrounds — can function as epistemic artefacts that operationalize lived experience and complement data-driven insights in intervention design. While personas have long mediated between design stakeholders [48], they are typically constructed by researchers or designers on behalf of users, a practice criticized for stereotyping and distance from lived realities, even when grounded in empirical data [8,38,55]. Building on participatory and value-sensitive design practices [21,49,53], our participatory personas were created by youth themselves, embedding first-person accounts of routines, challenges, and support needs, in direct response to limitations they identified in the data-driven personas developed by our research team. Our work examines how these participatory personas can operate as scaffolding devices that extend beyond user representation to serve as methodological tools, particularly to engage marginalized youth often overlooked in research alongside community members. Specifically, the current study personas were used to iteratively structure co-creation sessions, elicit stakeholder perspectives, and refine personalization priorities. In doing so, they bridged participatory design and technical development, enabling community perspectives to inform the training of LLMs for DMHIs. This repositions pesrsonas from representational artefacts that stand in for imagined "users" to expressive, person-centred artefacts that carry service users' voices directly into design decision-making.

**5.2 Practical Implications for Developing LLMs for DMHIs**

*5.2.1 For digital health research and development teams*

Our findings also have practical implications for digital health research and development teams interested in deploying LLMs in their work. Building on our theoretical claim that personalization can be framed as dialogic contextualization in lived experience, our findings illustrate how digital health tools, particularly those leveraging LLMs, can practically implement adaptation based on end-users' ongoing interactions. One example is the use of dialogic scaffolds such as interpretive questioning, intent clarification, autonomy support, and neutral tone, as identified in the current study. These techniques extend personalization beyond existing persuasive strategies and highlight concrete interactional features for future studies to evaluate in support of youth well-being. Our findings also suggest that initial onboarding in applications should set clear expectations of the LLM-based tool's role, limitations, and data handling in age-appropriate language, with explicit consent before use. Similarly, access to human support and offline resources must be visible and immediate to reinforce the role of chatbots as supplements, not substitutes for, human care.

While foundational prior research has provided valuable taxonomies of harmful behaviors and psychological risks associated with AI conversational agents [9, 25], our work extends these contributions by including exemplar, evidence-based dialogue extracts to fine-tune LLMs for health promotion while safeguarding. For LLM developers, a central takeaway from our study involves integrating the personalization priorities we identified into training data and system pipelines. However, current training approaches face important limitations in participatory engagement. Reliance on crowdsourced annotators, who may even use existing LLMs for annotation tasks, can limit the relevance and appropriateness of applications of LLMs for youth well-being [33, 72]. Therefore, lived experience experts who understand challenges specific to youth mental health, including youth themselves, are needed to transition applications of LLMs from generic support to adaptability to the needs of youth in particular.

Our findings outline several specific mechanisms to embed these stakeholder perspectives into LLM and DMHI system behavior. We offer guidance for how contextualization, which requires building pipelines that allow models to adjust responses based on patterns in user routines, the social or relational context in which challenges arise, and differences in literacy or communication style. Further, safeguards such as role limits, referral mechanisms, and data transparency identified in the present work can be treated as engineering requirements rather than add-ons, to shape dialogue and system responses from the outset. Dialogic scaffolds identified in our study can serve as templates for conversation design and as preference signals for reward models, ensuring that systems prioritize reflective, neutral, and autonomy-supportive responses rather than overly empathetic ones, as suggested by our study experts. In sum, the persona-driven insights and dialogue extracts derived in this study provide training material for supervised fine-tuning and hardcoding, enabling targeted and incremental alignment of LLM system behavior with co-created, community-informed preferences and safeguards [73].

*5.2.2 For public health policymakers*

Our work contributes to growing responses for investigating opportunities and solutions for responsible AI in mental health, particularly with youth, given significant mental health risks and digital engagement in this population globally [3, 31]. As this field develops rapidly, our study demonstrates the practical value of investing in community-based co-creation for stakeholder alignment and highlights structured methods for doing so. Community members and mental health professionals, such as youth, parents, and youth care workers, are critical stakeholders in the responsible design of well-being tools that leverage LLMs. In the current study, youth care workers demonstrated both willingness and motivation to be involved in shaping how AI systems can support youth mental well-being and prevent mental health challenges. Our study shows how participatory processes can offer a clear pathway for this involvement, enabling community stakeholders to help define the boundaries of digital health personalization, determine when and what specific referral to offline resources is appropriate, and clarify how digital tools can complement rather than replace real relationships and professional care. Further, while community members may be cautious about AI, when involved in collaborative design activities, they gain exposure to how LLMs work, how youth might use them, and what safeguards are needed for responsible integration into care, which can importantly assist them in making informed decisions about adopting such digital tools. Our findings thus highlight the practical value of investing in community-based co-creation for responsible innovation in digital mental health and health systems. Policy investment in such participatory infrastructures would enable sustained collaboration between practitioners, researchers, and communities, strengthening capacity to monitor, evaluate, and govern the integration of LLMs in community and care settings, helping ensure that emerging AI systems for youth mental health are equitable.

**5.3 Limitations and Future Directions**

A key limitation of the current scope is that it remains to be verified whether the expected LLM behaviors identified in this study can be internalized by LLMs after fine-tuning on the current study outputs; this marks a key avenue for future work. LLM alignment requires not only designing high-quality training examples but also evaluating whether these behaviors consistently emerge in model outputs. This highlights the need for further iterative evaluation loops between participatory design, model training, and real-world testing. Future work should therefore focus on empirical validation, testing how well the current alignment guidelines translate into observable model behaviors, and adapting training strategies accordingly.

Further, personas, even when created by end-users, can carry biases based on how participants frame experiences. Our participatory personas were generated directly by youth filling in semi-structured templates. While this approach was accessible for youth, it also produced static snapshots compared with "dynamic personas" in interaction with different scenarios, which would further enhance comprehensiveness of lived experience insights. This is why our current work emphasized using personas as artefacts for guiding dialogue between stakeholders during early stages of design, rather than as decision points and/or mechanisms for implementation in digital systems personalization. Further work should explore how participatory personas in combination with dynamic scenarios can be leveraged in mature stages of personalization and LLM fine-tuning, for example through fine-tuning of user and contextual profiling to augment information retrieval.

In addition, our study sample was relatively small and localized to the Dutch community youth care context, and youth participants were primarily older rather than younger adolescents. While not generalizable, our sample highlights perspectives underrepresented in mental health AI research and applications, specifically marginalized youth from low socioeconomic backgrounds in community settings, which has important implications for designing equitable DMHIs. Further, our triangulation of multiple community stakeholders' perspectives on personalization (youth care workers, youth, and parents) surfaced insights highly relevant to youth and community-centered design more broadly. However, while subthemes were consistent across stakeholders, we did not explicitly explore value tensions or contradictions between how stakeholder groups (youth, youth care workers, parents) perceived the different personas and persona creation processes; this was out of the current study scope, but remains critical for future applications of personas in LLM-alignment exercises. Further research is needed to refine and systematically outline person co-creation methods throughout LLM development stages, particularly to guide mature stages of deployment.

Finally, our study scope was limited to the general well-being and preventative health context, which focuses on prevention of severe mental health symptoms, given its significant role in public health outcomes. Future research is needed to extend the current findings to consider LLM personalization to support clinical health or specialist domains such as disease support. Engagement of specialist and clinical stakeholders in addition to patient communities will be crucial in such an endeavor. The current study methods can be adopted for further investigation with mental health counselors and psychiatrists to identify opportunities for integration into their work.

## 6 CONCLUSION

This work demonstrates how co-created dialogue strategies can guide personalization in LLMs for DMHIs supporting youth. We contribute a structured approach for embedding lived experience from community stakeholders into personalization design, shifting adaptive support beyond static tailoring toward interactive processes that foster reflection and autonomy. Future research should examine how these strategies can be integrated into LLM fine-tuning pipelines and evaluate their impact on mental well-being and quality-of-life outcomes, while further refining systematic participatory methods for the responsible development of AI in youth mental health.

## ACKNOWLEDGMENTS

We would like to thank Easha Jadnanansing, Sarah Duister, Kirmina Rezk, Neslihan Can, and Dimpy Gupta for their support in data collection and validation, without which this present work would not have been possible. We also thank Niko Vegt and Kayla Green for their insights during early stage co-creation sessions, and Atay Kozlovski for comments on an earlier version of this draft, which strengthened the current methods and interpretation of findings. Most importantly, we thank all youth and community participants for sharing their valuable time and expertise. This research is funded by the Health & Technology Flagship program, PROactive TEChnology-supported prevention and MEntal health in adolescence (PROTECt ME), as part of Convergence, the research alliance between Erasmus University Medical Center, Erasmus University Rotterdam, and Delft University of Technology. Additionally, KWG, CAF and MdR are supported through the 'High Tech for a Sustainable Future' capacity building program of the 4TU Federation in the Netherlands.

## STATEMENT ON RELATIONSHIP TO PRIOR WORK

This paper reports original work that has not been previously published. It builds on a study protocol preregistered prior to data collection on Open Science Framework (https://osf.io/khfxu/), which described our planned design and procedures but did not include findings, analysis, or theoretical framing.